\documentclass[aps,prd,superscriptaddress,notitlepage,showpacs,nofootinbib]{revtex4-1}
\usepackage{amsmath,mathrsfs}
\usepackage{epsfig}
\def\bea#1\eea{\begin{align}#1\end{align}}
\newcommand{\nnu}{\nonumber\\}
\newcommand{\bef}{\begin{figure}[htb]\centering}
\newcommand{\eef}{\end{figure}}

\begin{document}
\title{Quasi-parton distribution functions: a study in the diquark spectator model}

\author{Leonard Gamberg}
\email{lpg10@psu.edu}
\affiliation{Division of Science, 
                   Penn State Berks, 
                   Reading, PA 19610, USA}

\author{Zhong-Bo Kang}
\email{zkang@lanl.gov}
\affiliation{Theoretical Division, 
                   Los Alamos National Laboratory, 
                   Los Alamos, NM 87545, USA}

\author{Ivan Vitev}
\email{ivitev@lanl.gov}
\affiliation{Theoretical Division,
                   Los Alamos National Laboratory,
                   Los Alamos, NM 87545, USA}

\author{Hongxi Xing}
\email{hxing@lanl.gov}
\affiliation{Theoretical Division,
                   Los Alamos National Laboratory,
                   Los Alamos, NM 87545, USA}

\begin{abstract}
A set of quasi-parton distribution functions (quasi-PDFs) have been recently proposed by Ji. Defined as the matrix elements of equal-time spatial correlations, they can be computed on the lattice and should reduce to the standard PDFs when the proton momentum $P_z$ is very large. Since taking the $P_z\to \infty$ limit is not feasible in lattice simulations, it is essential to provide guidance for what values of $P_z$ the quasi-PDFs are good approximations of standard PDFs. Within the framework of the spectator diquark model, we  evaluate both the up and down quarks' quasi-PDFs and standard PDFs for all leading-twist distributions (unpolarized distribution $f_1$, helicity distribution $g_1$, and transversity distribution $h_1$). We find that, for intermediate parton momentum fractions $x$, quasi-PDFs are good approximations to standard PDFs (within $20-30\%$) when $P_z\gtrsim 1.5-2$ GeV. On the other hand, for large $x\sim 1$  much larger $P_z > 4$~GeV is necessary to obtain a satisfactory agreement between the two sets. We further test the Soffer positivity bound, and find that it does not hold in general for quasi-PDFs. 
\end{abstract}

\date{\today}

\maketitle

\section{Introduction}
Parton distribution functions (PDFs) are of fundamental importance to science. PDFs provide invaluable information on the proton's partonic structure~\cite{Boer:2011fh,Accardi:2012qut} and are essential ingredients in theoretical predictions and description of the data from high energy scattering experiments~\cite{Lai:2010vv,Martin:2009iq}. At the same time, the calculation of PDFs from first principles in QCD  remains a great challenge.  While their existence has been theoretically established through QCD factorization~\cite{Collins:1989gx,Collins:2011zzd}, PDFs are essentially non-perturbative quantities and thus, cannot be obtained using  perturbative QCD techniques. Thus far, PDF extraction  from the experimental data has relied on a global fitting procedure within the standard factorization framework~\cite{Collins:1989gx,Collins:2011zzd}.

In recent years, however,
evaluation of PDFs has been attempted in lattice QCD~\cite{Deka:2008xr,Alexandrou:2014yha,Hagler:2009ni,Musch:2011er}.
Since PDFs are defined as the non-local light-cone correlations which involve the real Minkowski time, the traditional lattice QCD approach does not allow one to compute the PDFs {\em directly}~\cite{Ji:2013dva}; one can only calculate the lower moments of the PDFs, which are matrix elements of local operators~\cite{Deka:2008xr,Alexandrou:2014yha}. Recently, new methods have been proposed \cite{Ji:2013dva,Ma:2014jla,Ji:2014gla} to evaluate  PDFs on the lattice in terms of so-called  quasi-PDFs,  introduced by Ji \cite{Ji:2013dva}, which are
 defined as matrix elements of equal-time spatial correlators. 
These quasi-PDFs can be computed directly on the lattice~\cite{Lin:2014zya,Alexandrou:2014pna} and should reduce to the standard PDFs when the proton's momentum $P_z\to \infty$~\cite{Ji:2014gla}. While in practice the proton momentum on the lattice can never become infinite, one can only  hopefully  access finite but large enough momenta on the lattice to carry out relevant QCD simulations. 
For more details on quasi-PDFs the reader is referred to~\cite{Ji:2013dva,Ma:2014jla,Xiong:2013bka}, where factorization theorems are derived to connect the quasi-PDFs at finite $P_z$ to the standard PDFs through calculable coefficient functions. 

In this paper, we take a slightly different approach. Within the framework of a spectator diquark model~\cite{Jakob:1997wg, Gamberg:2003ey, Gamberg:2007wm, Bacchetta:2008af, Kang:2010hg} we compute both the quasi-PDFs and the standard PDFs, to study/explore for what values of $P_z$ they are good approximations of each other. This can provide guidance for future lattice QCD calculations.  
At leading-twist, the state of quarks in the proton is characterized by three distinct distribution functions: $f_1(x)$ the unpolarized parton distribution, $g_1(x)$ the helicity distribution, and $h_1(x)$ the transversity distribution. We evaluate these leading-twist distributions for both quasi-PDFs and standard PDFs and  for  up and down quarks. First, we formally verify that all quasi-PDFs reduce to the standard PDFs when the proton momentum $P_z\to \infty$. At the same time, we find that for the intermediate $x$ region ($0.1\lesssim x\lesssim 0.4-0.5$), the quasi-PDFs approximate the standard PDFs to  within $20-30\%$ when $P_z\gtrsim 1.5-2$ GeV. However, we find that for  large $x\sim 1$, one has to go to much larger $P_z > 4$ GeV  to ensure that the quasi-PDFs approach the standard PDFs. We further test the Soffer positivity bound~\cite{Artru:2008cp,Bacchetta:1999kz} for the quasi-PDFs and find that, in general, the usual positivity bounds do not hold for quasi-PDFs. 

The rest of our paper is organized as follows. In Sec.~II, we provide a short overview of both the standard PDFs and quasi-PDFs. We introduce the notation, and present the operator definitions for $f_1(x), ~g_1(x),~h_1(x)$ and the corresponding quasi-PDFs $\tilde f_1(x, P_z), ~\tilde g_1(x, P_z), ~\tilde h_1(x, P_z)$. 
We further define the cut vertices for these distributions, which will be used in the model calculations. In Sec.~III, we provide the analytical calculations for both quasi-PDFs and the standard PDFs, within the diquark model
for  the  scalar and axial-vector spectators.  At the end of this section, we briefly discuss the Soffer bound using the analytic expressions. In Sec.~IV, we present our numerical studies.  We use previously fitted parameters for the spectator diquark model, which lead to reasonable standard PDFs, consistent with those extracted from the global analysis. We then study the behaviors of both quasi-PDFs and the standard PDFs for all  three leading-twist distributions, and for both up and down quarks. Based on these numerical studies, we estimate at what values of $P_z$ the quasi-PDFs are good approximations of the standard PDFs. We also test the Soffer bound numerically for the quasi-PDFs. We conclude our paper in Sec.~V.

\section{Standard PDFs and Quasi-PDFs: overview and definitions}
In this section we provide a short introduction and definitions for the standard PDFs at leading-twist: the unpolarized distribution $f_1(x)$, the helicity distribution $g_1(x)$, and the transversity distribution $h_1(x)$. We also discuss the corresponding quasi-PDFs, $\tilde f_1(x, P_z)$, $\tilde g_1(x, P_z)$, and $\tilde h_1(x, P_z)$.

We consider a nucleon of mass $M$ moving in the $z$-direction, with the momentum $P^\mu$ given by
\bea
P^\mu=(P_0, 0_\perp, P_z) \equiv [P^+, P^-, 0_\perp].
\eea
Here and throughout the paper we use $(v_0, v_\perp, v_z)$ and $[v^+, v^-, v_\perp]$ to represent Minkowski and light-cone components for any four-vector $v^\mu$ respectively, with light-cone variables $v^{\pm} = (v_0\pm v_z)/\sqrt{2}$. We thus have
\bea
P^- = \frac{M^2}{2P^+}, \qquad
P_0 = \sqrt{P_z^2 + M^2} \equiv P_z \delta,
\eea
where $\delta$ is given by
\bea
\delta=\sqrt{1+\frac{M^2}{P_z^2}}.
\label{eq:delta}
\eea
For the helicity distribution $g_1$  and the transversity distribution $h_1$ we also have to consider the nucleon with either longitudinal or transverse polarization. For  pure longitudinal polarization, $S_L^\mu$ and 
transverse polarization, $S_T^\mu$ 
the polarization vectors are, 
\bea
S_L^\mu =\frac{1}{M}\left(P_z, 0_\perp, P_0\right) \equiv \frac{1}{M}\left[P^+, -P^-, 0_\perp\right], \qquad 
S_T^\mu = (0, \vec{S}_\perp, 0) \equiv [0^+, 0^-, \vec{S}_\perp].
\label{eq:sl}
\eea
The polarization vectors satisfy the conditions  $P\cdot S_L = P\cdot S_T = 0$,  and $S_L^2 = -1$ and $S_T^2 = -\vec{S}_\perp^2 = -1$.

The three leading-twist standard collinear PDFs are defined on the light-cone with the following operator expressions~\cite{Brock:1993sz}
\bea
f_1(x) &= \int \frac{d\xi^-}{4\pi} e^{-i \xi^-k^+}  \langle P|\overline{\psi}(\xi^-) \gamma^+ U_n[\xi^-,0] \psi(0) |P\rangle,
\\
g_1(x) &=\int \frac{d\xi^-}{4\pi} e^{-i \xi^-k^+}  \langle PS|\overline{\psi}(\xi^-) \gamma^+\gamma_5 U_n[\xi^-,0] \psi(0) |PS\rangle,
\\
h_1(x) &=\int \frac{d\xi^-}{4\pi} e^{-i \xi^-k^+}  \langle PS|\overline{\psi}(\xi^-) \gamma^+\gamma_5 \gamma\cdot S_T U_n[\xi^-,0] \psi(0) |PS\rangle,
\eea   
with $x=k^+/P^+$. We define the light-cone vector $n^\mu = [0^+, 1^-, 0_\perp]$ with $n^2 = 0$ and $n\cdot v=v^+$ for any four-vector $v^\mu$, and  
the gauge link $U_n[\xi^-,0]$ along the light-cone direction specified by $n$ is given by
\bea
U_n[\xi^-,0] = \exp\left(-ig\int^{\xi^-}_0 d\eta^- A^+(\eta^-) \right).
\eea
On the other hand, the quasi-PDFs introduced by Ji \cite{Ji:2013dva} are equal-time spatial correlations along the $z$-direction, and have the following operator definitions
\bea
\tilde f_1(x, P_z) &= \int \frac{d\xi_z}{4\pi} e^{ -i \xi_z k_z}  \langle P|\overline{\psi}(\xi_z) \gamma_z U_{n_z}[\xi_z,0] \psi(0) |P\rangle,
\\
\tilde g_1(x, P_z) &=\int \frac{d\xi_z}{4\pi} e^{-i \xi_z k_z}  \langle PS|\overline{\psi}(\xi_z) \gamma_z\gamma_5 U_{n_z}[\xi_z,0] \psi(0) |PS\rangle,
\\
\tilde h_1(x, P_z) &=\int \frac{d\xi_z}{4\pi} e^{-i \xi_z k_z}  \langle PS|\overline{\psi}(\xi_z) \gamma_z \gamma_5 \gamma\cdot S_T U_{n_z}[\xi_z,0] \psi(0) |PS\rangle,
\eea
where $n_z^\mu = (0, 0_\perp, 1)$ with $n_z^2 = -1$ and $n_z \cdot v = -v_z$ for any four-vector $v^\mu$, where now the gauge link $U_{n_z}[\xi_z,0]$ is 
along the direction of $n_z$ and is  given by
\bea
U_{n_z}[\xi_z,0] = \exp\left(-ig\int^{\xi_z}_0 d\eta_z A_z (\eta_z)   \right).
\eea

The above collinear PDFs and quasi-PDFs can be represented by the cut forward scattering diagram in Fig.~\ref{fig:represent} with proper cut vertices~\cite{Kang:2008ey,Ma:2014jla}. 
\bef
\psfig{file=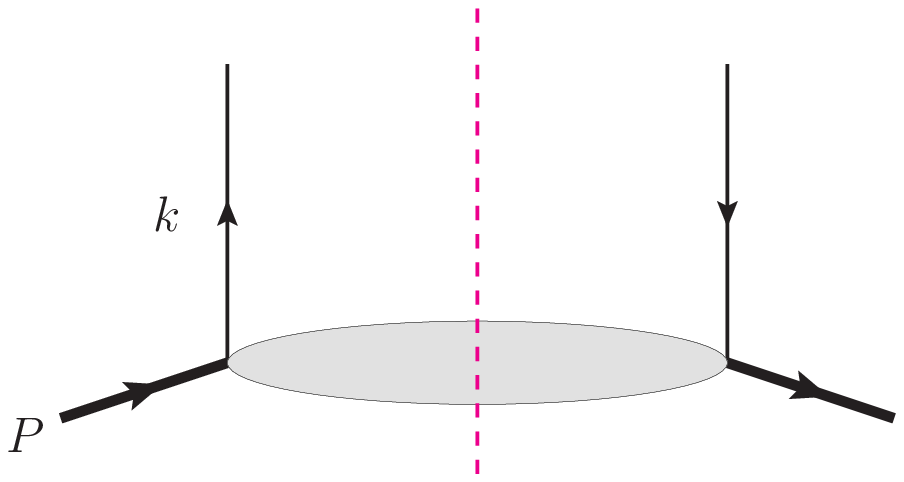, width=2in}
\caption{\scriptsize The generic Feynman diagram representation for the leading-twist PDFs and the corresponding quasi-PDFs.}
\label{fig:represent}
\eef
The  various leading-twist PDFs ($f_1$, $g_1$, and $h_1$) 
are characterized by the  cut vertices contracted with active partons in the diagram, and the cut vertices for the standard  PDFs are given by
\bea
f_1(x): \, \frac{\gamma^+}{2P^+} \delta\left(x-\frac{k^+}{P^+}\right),
 \quad g_1(x): \, \frac{\gamma^+\gamma_5}{2P^+} \delta\left(x-\frac{k^+}{P^+}\right),
 \quad h_1(x): \, \frac{\gamma^+\gamma_5\gamma\cdot S_T}{2P^+} \delta\left(x-\frac{k^+}{P^+}\right).
\label{eq:f1}
\eea
On the other hand, the corresponding cut vertices for the quasi-PDFs are given by
\bea
\tilde f_1(x, P_z): \, \frac{\gamma_z}{2P_z} \delta\left(x-\frac{k_z}{P_z}\right),
 \quad \tilde g_1(x, P_z): \, \frac{\gamma_z\gamma_5}{2P_z} \delta\left(x-\frac{k_z}{P_z}\right),
 \quad \tilde h_1(x, P_z): \, \frac{\gamma_z\gamma_5\gamma\cdot S_T}{2P_z} \delta\left(x-\frac{k_z}{P_z}\right).
\label{eq:qf1}
\eea
Given these  well-defined cut vertices, we will  calculate both the standard
and quasi-PDFs, which will be the main focus of the next section. 

\section{Standard PDFs and Quasi-PDFs in spectator diquark model}
In this section we first give a short overview of the spectator diquark model. We then present the analytical calculations for the standard PDFs and the quasi-PDFs, and discuss certain features and observations. 

\subsection{The spectator diquark model}
The spectator diquark model of the nucleon has been described in great detail~\cite{Jakob:1997wg, Gamberg:2007wm, Bacchetta:2008af, Kang:2010hg}. Here we present a brief overview. In the spectator diquark model, the PDFs, which are traces of the quark-quark correlation functions as defined in the last section, are evaluated in the spectator approximation. In this framework a sum over a complete set of intermediate on-shell states, 
 $\openone=\sum_{X}|X\rangle\langle X|$, is  inserted into the operator definition of PDFs, and truncated to single on-shell diquark spectator states with $X$ being 
either spin 0 (scalar diquark) or spin 1 (axial-vector diquark). The quark-quark correlation function is then obtained as the cut tree level amplitude for nucleon $N\rightarrow q\, + \, X$ where $X={\{s,a\}}$. With such an approximation, the nucleon is composed of a constituent quark of mass $m$ and a spectator scalar (axial-vector) diquark with mass $M_s$ ($M_a$). 
\bef
\psfig{file=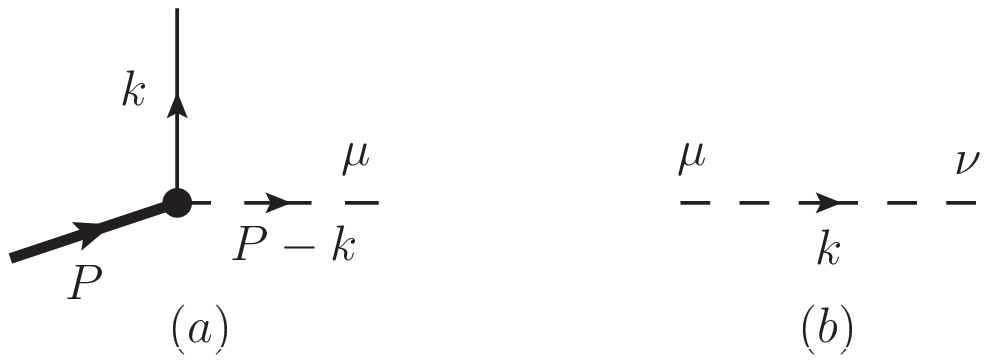, width=3in}
\caption{\scriptsize Feynman rules in the spectator diquark model: (a) vertex representing the interaction between the quark, the nucleon, and the diquark, (b) the diquark propagator.}
\label{fig:feynmanrule}
\eef
The interaction between the nucleon, the quark, and the diquark is given by the following Feynman rules for the vertex in Fig.~\ref{fig:feynmanrule}(a),
\bea
\mbox{scalar diquark:}  \,\,  ig_s \mathcal{I}_s(k^2)\, ,
\qquad \mbox{axial-vector diquark:}\,\,  i\frac{g_a}{\sqrt{2}} \gamma^\mu \gamma_5 \mathcal{I}_a(k^2),
\label{eq:axial-vertex}
\eea
where following~\cite{Gamberg:2007wm,Bacchetta:2008af, Kang:2010hg}, we have introduced suitable form factors $\mathcal{I}_{s,a}(k^2)$ as a function of $k^2$ - the invariant mass of the constituent quark. For our numerical calculations below, 
we adopt the fitted parameters in~\cite{Bacchetta:2008af} and use
 the  dipolar form factors,
\bea
\mathcal{I}_s(k^2) = \frac{k^2-m^2}{\left(k^2 - \Lambda_s^2\right)^2},
\qquad
\mathcal{I}_a(k^2) = \frac{k^2-m^2}{\left(k^2 - \Lambda_a^2\right)^2},
\label{eq:form}
\eea
where $\Lambda_{s,a}$ are the appropriate cutoffs, to be considered as free parameters of the model together with the diquark masses $M_{s,a}$, and the couplings $g_{s,a}$. Further, 
 the propagators of the scalar diquark and the axial-vector diquark as shown in Fig.~\ref{fig:feynmanrule}(b) are given by the expressions, 
\bea
\mbox{scalar diquark:} \,\,  \frac{i}{k^2-M_s^2},
\qquad \mbox{axial-vector diquark:} \,\, \frac{i}{k^2-M_a^2}d^{\mu\nu}(k, n),
\label{eq:axial-propagator}
\eea
where for the standard light-cone PDFs with $n^2=0$, we have~\cite{Bacchetta:2008af, Kang:2010hg}
\bea
d^{\mu\nu}(k, n)  = -g^{\mu\nu} + \frac{n^\mu k^\nu + n^\nu k^\mu}{n\cdot k} - \frac{k^2 n^\mu n^\nu}{\left(n\cdot k\right)^2},
\label{eq:axial-pol}
\eea
which satisfies $n_\mu d^{\mu\nu}(k, n) = k_\mu d^{\mu\nu}(k, n) = 0$. On the other hand, for the quasi-PDFs, since $n_z^2 =-1\neq 0$, we have a slightly different form for the polarization tensor $d^{\mu\nu}$ as
\bea
d^{\mu\nu}(k, n_z) =  -g^{\mu\nu} + \frac{n_z\cdot k}{\left(n_z\cdot k\right)^2 - n_z^2 k^2} 
\left(n_z^\mu k^\nu + n_z^\nu k^\mu\right)
- \frac{1}{\left(n_z\cdot k\right)^2 - n_z^2 k^2} \left(k^2 n_z^\mu n_z^\nu + n_z^2 k^\mu k^\nu
\right),
\label{eq:axial-quasi-pol}
\eea
which also satisfies $n_{z\mu} d^{\mu\nu}(k, n_z) = k_{\mu} d^{\mu\nu}(k, n_z) = 0$.

\subsection{Standard PDFs and quasi-PDFs in the diquark model}
In the lowest order calculation of the spectator diquark model, the leading-twist standard PDFs (or quasi-PDFs) are calculated from the Feynman diagram shown in Fig.~\ref{fig:diquark}. With the cut vertices in Eqs.~(\ref{eq:f1}) 
and (\ref{eq:qf1}), as well as the Feynman rules presented above, the calculation is straightforward. Below we demonstrate how to derive $f_1(x)$ and the unpolarized quasi-PDF $\tilde f_1(x, P_z)$ for both the scalar and axial-vector diquark as an example, and provide the  final results for $g_1(x)$  and $\tilde g_1(x, P_z)$,  and $h_1(x)$  and $\tilde h_1(x, P_z)$. Our results for the standard PDFs $f_1, ~g_1$ and $h_1$ are consistent with those in~\cite{Bacchetta:2008af}. We  present them here in order to compare  with the quasi-PDFs. 
\bef
\psfig{file=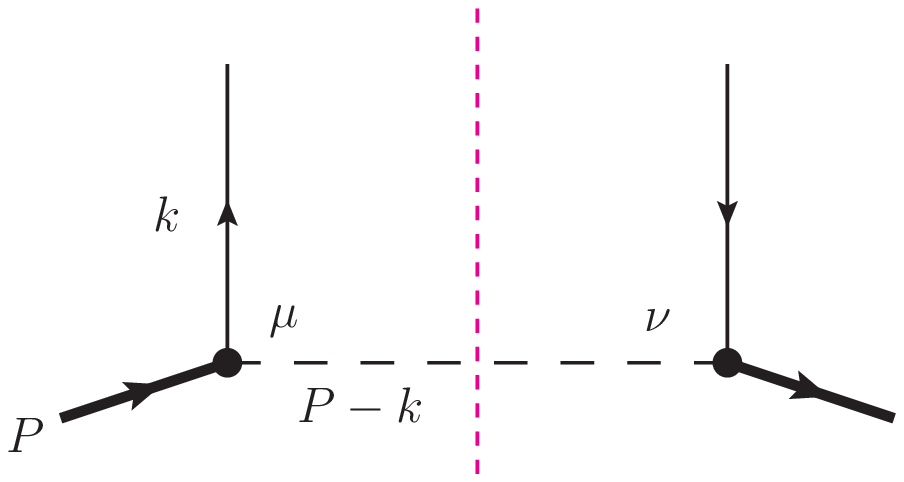, width=2in}
\caption{\scriptsize The lowest order Feynman diagram for the leading-twist standard PDFs (or quasi-PDFs) in the spectator diquark model.}
\label{fig:diquark}
\eef

\subsubsection{Unpolarized distributions: $f_1$ and $\tilde f_1$}
In the scalar diquark model~\cite{Jakob:1997wg}, $f_1^s(x, k_\perp^2)$ can be derived from Fig.~\ref{fig:diquark} and is given by
\bea
f_1^s(x, k_\perp^2) =& g_s^2 \int \frac{dk^+ dk^-}{(2\pi)^4} \frac{1}{2P^+} \delta\left(x - \frac{k^+}{P^+}\right){\rm Tr}\left[\gamma\cdot n \left(\gamma\cdot k+m \right) \frac{1}{2}\left(\gamma\cdot P+M\right)\left(\gamma\cdot k+m\right) \right]
\nnu
&\times
\frac{1}{(k^2-m^2)^2} 2\pi \delta\left(\left(P-k\right)^2 - M_s^2\right) \left[\mathcal{I}_s(k^2)\right]^2,
\eea
where the superscript ``$s$'' in $f_1^s$ indicates that the diquark is a scalar, and $k_\perp$ is the quark transverse momentum~\cite{Bacchetta:2004jz,Bacchetta:2006tn}. To proceed, we write 
\bea
\delta\left(\left(P-k\right)^2 - M_s^2\right) &= \delta\left(2(P^+ - k^+)(P^- - k^-) - k_\perp^2 - M_s^2 \right)
= \frac{1}{2(1-x)P^+} \delta\left(k^- - \frac{M^2}{2P^+} + \frac{k_\perp^2 + M_s^2}{2(1-x)P^+} \right),
\eea
which can then be used to integrate over $dk^-$. On the other hand, $\delta(x-k^+/P^+)$ can be used to integrate over $dk^+$. Eventually we obtain
\bea
f_1^s(x,k_{\perp}^2)&=\frac{g_s^2}{(2\pi)^3}\frac{(1-x)[k_{\perp}^2+(m+xM)^2]}
{2\left[k_{\perp}^2+xM_s^2-x(1-x)M^2+(1-x)m^2\right]^2} \left[\mathcal{I}_s(k^2)\right]^2,
\label{eq:truef1}
\eea
where the invariant mass $k^2$ is given by
\bea
k^2= - \frac{1}{1-x} \left[k_\perp^2+x M_s^2 - x(1-x)M^2 \right].
\label{eq:k2}
\eea

Motivated by the definition of the cut vertices for the quasi-PDFs in Eqs.~(\ref{eq:qf1}) and Fig.~\ref{fig:diquark}, 
we  write the quasi-PDF $\tilde f_1^s(x, k_\perp^2, P_z)$ for the scalar diquark case as
\bea
\tilde f_1^s(x, k_\perp^2, P_z) =& - g_s^2 \int \frac{dk_0 dk_z}{(2\pi)^4} \frac{1}{2P_z} \delta\left(x - \frac{k_z}{P_z}\right){\rm Tr}\left[\gamma\cdot n_z \left(\gamma\cdot k+m \right) \frac{1}{2}\left(\gamma\cdot P+M\right)\left(\gamma\cdot k+m\right) \right]
\nnu
&\times
\frac{1}{(k^2-m^2)^2} 2\pi \delta\left(\left(P-k\right)^2 - M_s^2\right) \left[\mathcal{I}_s(k^2)\right]^2,
\eea
where we have used $\gamma_z = -\gamma\cdot n_z$. Now the on-shell condition for the scalar diquark can be written as
\bea
\delta\left(\left(P-k\right)^2-M_s^2\right) &= \delta \left(\left(P_0-k_0\right)^2 - \left(P_z-k_z\right)^2 - k_\perp^2 - M_s^2 \right)
= \frac{1}{2\left(P_0 - k_0\right)}\delta \left(P_0 - k_0 - \lambda\right),
\label{eq:onshell}
\eea
where $\lambda$ is given by
\bea
\lambda \equiv \sqrt{(1-x)^2P_z^2 + k_\perp^2 + M_s^2} = (1-x) P_z \rho_s,
\eea
with the following expression for $\rho_s$ 
\bea
\rho_s \equiv \sqrt{1+\frac{k_{\perp}^2+M_s^2}{(1-x)^2P_z^2}}.
\label{eq:rho}
\eea
Now,  we  integrate over $dk_0$ with the help of $\delta\left(P_0 - k_0 - \lambda\right)$,  
and also use $\delta(x - k_z/P_z)$ to integrate over $dk_z$, setting $k_z = x P_z$. 
Finally, the corresponding quasi-PDF $\tilde f_1^s(x, k_\perp^2, P_z)$ is given by
\bea
{\tilde f}_1^s(x,k_{\perp}^2,P_z)&=\frac{g_s^2}{(2\pi)^3} \frac{\mathcal{F}_s}
{\mathcal{D}_s} \left[\mathcal{I}_s(k^2)\right]^2,
\label{eq:tilde-f1}
\eea
where the factors $\mathcal{F}_s$ and $\mathcal{D}_s$ are 
is defined as
\bea
\mathcal{F}_s \equiv & \, (2x-1)M^2+2x M m-M_s^2+m^2-2(1-x)^2(1-\rho_s\delta)P_z^2\, ,\\
\mathcal{D}_s \equiv & \, 2\rho_s(1-x)\left[2(1-x)(1-\rho_s\delta)P_z^2+M^2+M_s^2-m^2\right]^2\,  ,
\eea
and  $\rho_s$ is given by Eq.~\eqref{eq:rho} and
 $\delta$, Eq.~\eqref{eq:delta}. For the quasi-PDFs,  $k^2$ is
\bea
k^2 = 2(1-x)(1-\rho_s\delta)P_z^2+ M^2+M_s^2\, .
\label{eq:qk2}
\eea

We now study what happens to the quasi-PDF $\tilde f_1^s(x, k_\perp^2, P_z)$ 
in the limit  of $P_z\to \infty$. 
Approximating  $\rho_s$ and $\delta$ to $\mathcal{O}(M^2/P_z^2)$
\bea
\rho_s & \approx 1+ \frac{k_{\perp}^2+M_s^2}{2(1-x)^2P_z^2},
\qquad \delta   \approx 1+\frac{M^2}{2P_z^2},
\label{eq:rho_s}
\eea
where the quantity becomes, 
\bea
(1-\rho_s\delta) P_z^2 \approx - \frac{k_{\perp}^2+M_s^2}{2(1-x)^2} - \frac{M^2}{2},
\eea
and substituting this expression  into both Eqs.~\eqref{eq:tilde-f1} and \eqref{eq:qk2}, we  find that
\bea
\tilde f_1^s(x, k_\perp^2, P_z\to \infty) = f_1^s(x, k_\perp^2)\, .
\eea
Thus, the quasi-PDF reduces to the standard PDF $f_1^s(x, k_\perp^2)$ as  $P_z\to\infty$ limit~\footnote{Though obvious, it is worthwhile emphasizing that this conclusion is independent of whether or not one uses the form factor $\mathcal{I}_s(k^2)$ in the spectator diquark model.}. This simply verifies the leading order matching calculations carried out in \cite{Ji:2013dva,Ma:2014jla}. 

The approximation for $\rho_s$ used in Eq.~\eqref{eq:rho_s}  seems quite reasonable. However, it is important to emphasize that such an approximation only holds when $(1-x)^2 \sim \mathcal O(1)$. When we are studying the quasi-PDFs in the very large $x\sim 1$ region, the large $P_z$ expansion used for $\rho_s$
 breaks down, in which case the quasi-PDFs can deviate substantially 
from the standard PDFs. Such a breakdown is directly related to the existence of the factor $(1-x)^2P_z^2$ in our calculation, which is traced back to the on-shell condition of the diquark in Eq.~\eqref{eq:onshell}. Since such an on-shell condition is fairly generic~\cite{Xiong:2013bka}, 
we expect that it will be quite difficult for the quasi-PDFs to approach the standard PDFs in the large $x\sim 1$ region. In this case, one has to boost the proton to much larger $P_z$.  We will further illustrate this point in our numerical studies in the next section.

With the dipolar form factor $\mathcal{I}_s(k^2)$ given in Eq.~\eqref{eq:form}, one can further integrate $f_1^s(x, k_\perp^2)$ over $k_\perp^2$ to obtain the collinear distribution $f_1^s(x)$ as 
\bea
f_1^s(x) = \int d^2k_\perp f_1^s(x,k_{\perp}^2) =  2\pi\int_0^{\infty} dk_\perp k_\perp f_1^s(x,k_{\perp}^2),
\eea
from which we obtain
\bea
f_1^s(x) = \frac{g_s^2}{(2\pi)^2}\frac{\left[2(m+x M)^2 + L_s^2(\Lambda_s^2)\right] (1-x)^3}{24L_s^6(\Lambda_s^2)},
\eea
with $L_s^2(\Lambda_s^2)$ defined as
\bea
L_s^2(\Lambda_s^2) \equiv xM_s^2 + (1-x) \Lambda_s^2 - x(1-x) M^2.
\eea
Now  let us consider  the quasi-PDF $\tilde f_1^s(x, P_z)$. We have
\bea
\tilde f_1^s(x,P_z) = \int d^2k_\perp \tilde f_1^s(x,k_{\perp}^2, P_z) =  2\pi\int_0^{\infty} dk_\perp k_\perp \tilde f_1^s(x,k_{\perp}^2, P_z),
\eea
with $\tilde f_1^s(x,k_{\perp}^2, P_z)$ given by Eq.~\eqref{eq:tilde-f1}. Because of the complicated functional form for $\tilde f_1^s(x,k_{\perp}^2, P_z)$, we are not able to obtain a simple analytical expression for the collinear quasi-PDF $\tilde f_1^s(x, P_z)$ and will only present the numerical studies for the collinear quasi-PDFs in the next section. Here, it is important to 
emphasize that, since in the limit of $P_z\to \infty$, $\tilde f_1^s(x, k_\perp^2, P_z)$ reduces to $f_1^s(x, k_\perp^2)$ as we have shown above, the collinear counter-part $\tilde f_1^s(x, P_z)$ also reduces to the standard collinear PDF $f_1^s(x)$. 

Let us now turn to the calculation of both $f_1$ and $\tilde f_1$ for the axial-vector diquark case. The calculations are very similar to those above. The differences are: (a) the nucleon-quark-diquark vertex is now given by Eq.~\eqref{eq:axial-vertex},  (b) the diquark propagator is now given by Eq.~\eqref{eq:axial-propagator}, and  (c) in the standard PDF calculation, one uses  the polarization sum in Eq.~\eqref{eq:axial-pol} for the axial-vector diquark, while in the quasi-PDF calculation, the polarization sum,  Eq.~\eqref{eq:axial-quasi-pol} 
is used, where since now $n_z^2\neq 0$. We now have for the standard PDF
\bea
f_1^a(x,k_{\perp}^2) & = \frac{g_a^2}{(2\pi)^3} \frac{(1+x^2)k_{\perp}^2+(1-x)^2(m+xM)^2}
{2(1-x)\left[k_{\perp}^2+xM_a^2-x(1-x)M^2+(1-x)m^2\right]^2} \left[\mathcal{I}_a(k^2)\right]^2,
\label{eq:tf1a}
\\
f_1^a(x) &=  \frac{g_a^2}{(2\pi)^2} \frac{\left[2(m+x M)^2(1-x) + (1+x^2) L_a^2(\Lambda_a^2)\right] (1-x)}{24L_a^6(\Lambda_a^2)},
\eea
where the superscript ``$a$'' represents the axial-vector diquark, and $L_a^2(\Lambda_a^2)$ is given by
\bea
L_a^2(\Lambda_a^2) \equiv xM_a^2 + (1-x) \Lambda_a^2 - x(1-x) M^2.
\eea
At the same time, the quasi-PDFs for the axial-vector diquark case are 
\bea
{\tilde f}_1^a(x,k_{\perp}^2,P_z)=\frac{g_a^2}{(2\pi)^3} \frac{\mathcal{F}_a}{\mathcal{F}_b} \left[\mathcal{I}_a(k^2)\right]^2,
\qquad
\tilde f_1^a(x,P_z)  = \int d^2k_\perp \tilde f_1^a(x,k_{\perp}^2, P_z),
\eea
where the factors $\mathcal{F}_a$ and $\mathcal{F}_b$ are given by
\bea
\mathcal{F}_a \equiv & \left[M_a^2+(1-x)^2P_z^2\right]\left[2(1-x)^2(1-\rho_a\delta)P_z^2+M^2-2xmM-m^2+M_a^2\right]
\nnu
&+2x(1-x)^2 P_z^4(1-\rho_a^2\delta^2)+2xM_a^2P_z^2,
\\
\mathcal{F}_b \equiv  & -2\rho_a(1-x) \left[M_a^2+(1-x)^2P_z^2\right]\left[2(1-x)(1-\rho_a\delta)P_z^2+M^2+M_a^2-m^2\right]^2,
\eea
with $\rho_a$ given by
\bea
\rho_a &= \sqrt{1+\frac{k_{\perp}^2+M_a^2}{(1-x)^2P_z^2}} \approx 1+ \frac{k_{\perp}^2+M_a^2}{2(1-x)^2P_z^2}\, ,\quad \mbox{as $P_z\to \infty$}.
\label{eq:rho_a}
\eea
Using the expansions of $\rho_a$ and $\delta$ in the limit  $P_z\to \infty$, one  finds that 
\bea
{\tilde f}_1^a(x,k_{\perp}^2,P_z\to \infty) = f_1^a(x, k_\perp^2),
\eea
as given by Eq.~\eqref{eq:tf1a}.  Further, 
since in the limit of $P_z\to \infty$, $\tilde f_1^a(x, k_\perp^2, P_z)$ reduces to $f_1^a(x, k_\perp^2)$, again, the collinear counterpart: $\tilde f_1^a(x, P_z\to\infty) = f_1^a(x)$.  We thus demonstrate the matching of the quasi-PDFs to the standard PDFs for the axial-vector diquark case. 

\subsubsection{Helicity distributions: $g_1$ and $\tilde g_1$}
Following the same approach as in the previous section, we now present the results for the standard helicity distribution, $g_1(x)$,  and the quasi-helicity distribution, $\tilde g_1(x, P_z)$. Using the cut vertices for both the helicity distribution $g_1$ and quasi-helicity distribution $\tilde{g}_1$  given in Eqs.~\eqref{eq:f1} and \eqref{eq:qf1}, and in the expression Eq.~\eqref{eq:sl} for the longitudinal polarization vector of the nucleon $S_L^\mu$, the calculation is straightforward. The final results for the standard helicity distribution are given by
\bea
g_1^s(x,k_{\perp}^2)&=\frac{g_s^2}{(2\pi)^3} \frac{(1-x)[-k_{\perp}^2+(m+xM)^2]}
{2\left[k_{\perp}^2+xM_s^2-x(1-x)M^2+(1-x)m^2\right]^2} \left[\mathcal{I}_s(k^2)\right]^2,
\\
g_1^a(x,k_{\perp}^2)&=\frac{g_a^2}{(2\pi)^3} \frac{(1+x^2)k_{\perp}^2-(1-x)^2(m+xM)^2}
{2(1-x)\left[k_{\perp}^2+xM_a^2-x(1-x)M^2+(1-x)m^2\right]^2} \left[\mathcal{I}_a(k^2)\right]^2.
\eea

Carrying out a similar analysis as for the unpolarized quasi-PDFs,  the  quasi-helicity distributions are given by
\bea
{\tilde g}_1^s(x,k_{\perp}^2,P_z)&=\frac{g_s^2}{(2\pi)^3} \frac{\mathcal{G}_s}
{\mathcal{D}_s}  \left[\mathcal{I}_s(k^2)\right]^2,
\quad\quad {\tilde g}_1^a(x,k_{\perp}^2,P_z)=\frac{g_a^2}{(2\pi)^3} \frac{\mathcal{G}_a}{\mathcal{G}_b}  
\left[\mathcal{I}_a(k^2)\right]^2,
\label{eq:quasi_hel_TMD}
\eea
where the factors $\mathcal{G}_s$, $\mathcal{G}_a$, and $\mathcal{G}_b$ are given by
\bea
\mathcal{G}_s\equiv & 2(1-x)\rho_s P_z^2\left[(x-\delta^2)M+(1-\delta^2)m\right]+\delta M\left[(M+m)^2+M_s^2+2(1-x)^2P_z^2\right]\\
\mathcal{G}_a \equiv & M \left[M_a^2+(1-x)^2P_z^2\right]\left[\delta\left(M^2+m^2+M_a^2+2(x^2-x+1)P_z^2\right)
+2\rho_a(1-x)(x-\delta^2)P_z^2\right] 
\nnu
&-2x(1-x)^2\rho_a^2\delta M P_z^4 + 2(1-x)\rho_a(1-\delta^2)mP_z^2\left[M_a^2+(1-x)(1-x-\rho_a\delta)P_z^2\right],
\\
\mathcal{G}_b \equiv & -2\rho_a(1-x)M\left[M_a^2+(1-x)^2P_z^2\right]\left[2(1-x)(1-\rho_a\delta)P_z^2+M^2+M_a^2-m^2\right]^2.
\eea
The collinear helicity distributions are given by
\bea
g_1^s(x)  &= \int d^2k_\perp g_1^s(x,k_{\perp}^2) = \frac{g_s^2}{(2\pi)^2} \frac{\left[2(m+x M)^2 - L_s^2(\Lambda_s^2)\right] (1-x)^3}{24L_s^6(\Lambda_s^2)},
\\
g_1^a(x) &= \int d^2k_\perp g_1^a(x,k_{\perp}^2) = - \frac{g_a^2}{(2\pi)^2} \frac{\left[2(m+x M)^2(1-x) - (1+x^2)L_s^2(\Lambda_s^2)\right] (1-x)}{24L_a^6(\Lambda_a^2)}.
\eea
Using Eq.~\eqref{eq:quasi_hel_TMD},   the collinear quasi-helicity distributions are 
\bea
\tilde g_1^{s,a}(x, P_z)  &= \int d^2k_\perp \tilde g_1^{s,a}(x,k_{\perp}^2, P_z).
\eea
Using the expansions of $\rho_s$, $\rho_a$, and $\delta$ at large $P_z\to \infty$ in Eqs.~\eqref{eq:rho_s}, \eqref{eq:rho_a}, we can easily show that the quasi-helicity distribution functions reduce to the standard helicity distributions at $P_z\to\infty$,
\bea
\tilde g_1^{s,a}(x,k_{\perp}^2, P_z\to \infty) = g_1^{s,a}(x,k_{\perp}^2)\,  ,
\qquad
\tilde g_1^{s,a}(x, P_z\to \infty) = g_1^{s,a}(x).
\eea

\subsubsection{Transversity distributions: $h_1$ and $\tilde h_1$}
Here  we present the results for the transversity distributions. We have the following final results for the standard transversity distributions,
\bea
h_1^s(x,k_{\perp}^2)&=\frac{g_s^2}{(2\pi)^3} \frac{(1-x)(m+xM)^2}
{2\left[k_{\perp}^2+xM_s^2-x(1-x)M^2+(1-x)m^2\right]^2}  \left[\mathcal{I}_s(k^2)\right]^2,
\\
h_1^a(x,k_{\perp}^2)&=\frac{g_a^2}{(2\pi)^3} \frac{xk_{\perp}^2}
{(1-x)\left[k_{\perp}^2+xM_a^2-x(1-x)M^2+(1-x)m^2\right]^2}   \left[\mathcal{I}_a(k^2)\right]^2,
\eea
 and for the quasi-transversity distributions, 
\bea
{\tilde h}_1^s(x,k_{\perp}^2,P_z)&=\frac{g_s^2}{(2\pi)^3} \frac{\mathcal{H}_s}
{\mathcal{D}_s}\left[\mathcal{I}_s(k^2)\right]^2,\quad\quad
{\tilde h}_1^a(x,k_{\perp}^2,P_z)=\frac{g_a^2}{(2\pi)^3} \frac{\mathcal{H}_a}{\mathcal{H}_b} 
\left[\mathcal{I}_a(k^2)\right]^2,
\eea
where the factors $\mathcal{H}_s$,  $\mathcal{H}_a$, and $\mathcal{H}_b$ have the following forms
\bea
\mathcal{H}_s &\equiv k_{\perp}^2+(1-x)^2M^2-(m+xM)^2+M_s^2+2(1-x)^2(1-\rho_s\delta)P_z^2
\\
\mathcal{H}_a &\equiv k_{\perp}^2\left[-(1-x)^2M^2+(m+xM)^2+M_a^2\right]+4x(M^2+P_z^2)
\left[M_a^2+(1-x)^2P_z^2\right] -4x(1-x^2)\rho_a^2\delta^2P_z^4,
\\
\mathcal{H}_b &\equiv -4\rho_a(1-x)\left[M_a^2+(1-x)^2P_z^2\right]\left[2(1-x)(1-\rho_a\delta)P_z^2+M^2+M_a^2-m^2\right]^2.
\eea
The collinear transversity distributions are given by
\bea
h_1^s(x)  &= \int d^2k_\perp h_1^s(x,k_{\perp}^2) = \frac{g_s^2}{(2\pi)^2} \frac{(m+x M)^2 (1-x)^3}{12L_s^6(\Lambda_s^2)},
\\
h_1^a(x)  &= \int d^2k_\perp h_1^a(x,k_{\perp}^2) = -\frac{g_a^2}{(2\pi)^2} \frac{x (1-x)}{12L_a^4(\Lambda_a^2)},
\eea
while the collinear quasi-transversity distributions $\tilde h_1^{s,a}(x, P_z)$ are given by
\bea
\tilde h_1^{s,a}(x, P_z)  &= \int d^2k_\perp \tilde h_1^{s,a}(x,k_{\perp}^2, P_z).
\eea
One can also easily show the quasi-transversity distributions reduce to the standard transversity distributions,
\bea
\tilde h_1^{s,a}(x,k_{\perp}^2, P_z\to \infty) = h_1^{s,a}(x,k_{\perp}^2)\,  ,
\qquad 
\tilde h_1^{s,a}(x, P_z\to \infty) = h_1^{s,a}(x).
\eea

\subsubsection{Positivity bound: Soffer inequality}
For the standard PDFs, there are certain positivity bounds among them~\cite{Artru:2008cp,Bacchetta:1999kz,Kang:2011qz,Kang:2010fu}. For the leading-twist standard collinear PDFs, the Soffer inequality~\cite{Soffer:1994ww} gives a relation between the collinear unpolarized distribution $f_1(x)$, the helicity distribution $g_1(x)$, and the transversity distribution $h_1(x)$ for each flavor, 
\bea
|h_1(x)| \leq \frac{1}{2}\left(f_1(x) + g_1(x) \right).
\label{eq:soffer}
\eea
It is easy to verify that for the scalar diquark case
\bea
h_1^s(x) = \frac{1}{2}\left(f_1^s(x) + g_1^s(x) \right) =  \frac{g_s^2}{(2\pi)^2} \frac{(m+x M)^2 (1-x)^3}{12L_s^6(\Lambda_s^2)},
\eea
that is, they  saturate the Soffer bound. On the other hand, for the axial-vector diquark case, we have
\bea
\frac{1}{2}\left(f_1^a(x) + g_1^a(x) \right) - |h_1^a(x)| = \frac{g_a^2}{(2\pi)^2} \frac{(1-x)^3}{24L_a^4(\Lambda_a^2)} \geq 0,
\eea
for $0\leq x\leq 1$, thus  satisfying the Soffer bound. It is
 then  very interesting to test the Soffer bound for the quasi-PDFs; that is, to test whether we have
\bea
|h_1^{s,a}(x, P_z)| \stackrel{?}{\leq} \frac{1}{2}\left(f_1^{s,a}(x, P_z) + g_1^{s,a}(x, P_z) \right).
\eea
Since we do not have simple analytical expressions for the quasi-PDFs, we will
numerically test the Soffer bound for the quasi-PDFs  in the next section. 

\section{Numerical studies for standard PDFs and Quasi-PDFs}

Following Ref.~\cite{Bacchetta:2008af}, the $u$-quark and $d$-quark unpolarized PDFs $f_1^{u,d}$ can be written as
\bea 
f_1^u &= c_s^2 f_1^{u(s)} + c_a^2 f_1^{u(a)},
\\
f_1^d & = c_a'^{2} f_1^{d(a')},
\eea
that is, the $u$-quark receives contributions from both scalar and axial-vector diquark, while the $d$-quark only has the axial-vector diquark contribution. Here the superscript ``$s$'' represents the scalar diquark contribution, ``$a$'' corresponds to the axial-vector diquark which has isospin 0 (isoscalar $ud$-like system), and ``$a'$'' denotes the axial-vector diquark contribution which has isospin 1 (isovector $uu$-like system). Thus, we have the following 9 model parameters: $c_{s,a}$, $c_a'$, $M_{s,a}$, $M_a'$, $\Lambda_{s,a}$, and $\Lambda_a'$, as well as three couplings $g_s$, $g_a$, and $g_a'$. We use the same method specified in \cite{Bacchetta:2008af} to fix these three couplings: 
\bea
\pi \int_0^1 dx \int_0^{\infty} dk_\perp^2 f_1^{q(X)}(x, k_\perp^2) = 1,
\eea
with $X=s, a, a'$. On the other hand, the other 9 model parameters are fixed through a global fitting of both $f_1^u(x)$, $f_1^d(x)$ at factorization scale $\mu^2=0.30 {\rm ~GeV}^2$ with ZEUS2002 PDFs~\cite{Chekanov:2002pv} and $g_1^u(x)$, $g_1^d(x)$ at $\mu^2=0.26 {\rm ~GeV}^2$  with GRSV2000~\cite{Gluck:2000dy} at leading order in~\cite{Bacchetta:2008af};  the fit is satisfactory and gives consistent shape and size of the standard PDFs. In the following, we simply use these fitted parameters in our numerical study: specifically we use the parameters in Table I of Ref.~\cite{Bacchetta:2008af}. Once these parameters are fixed, we have expressions for both standard PDFs and the quasi-PDFs, and  are able to study them numerically to see if there are any interesting features or insights one might acquire. For example,  what values of $P_z$ quasi-PDFs are good approximations of standard PDFs, and whether the positivity bounds are satisfied for quasi-PDFs?  
\bef
\psfig{file=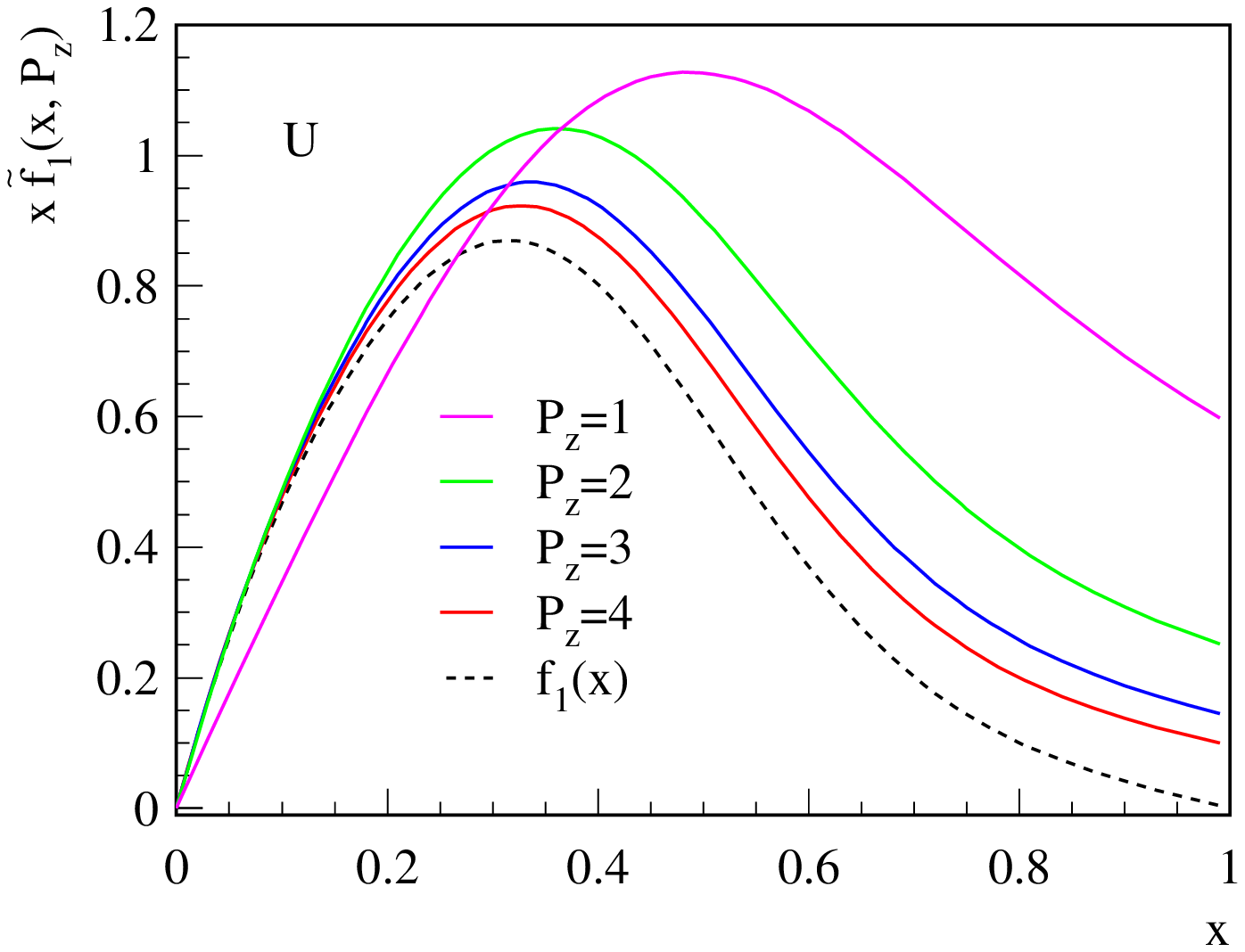, width=2.8in}
\hskip 0.4in
\psfig{file=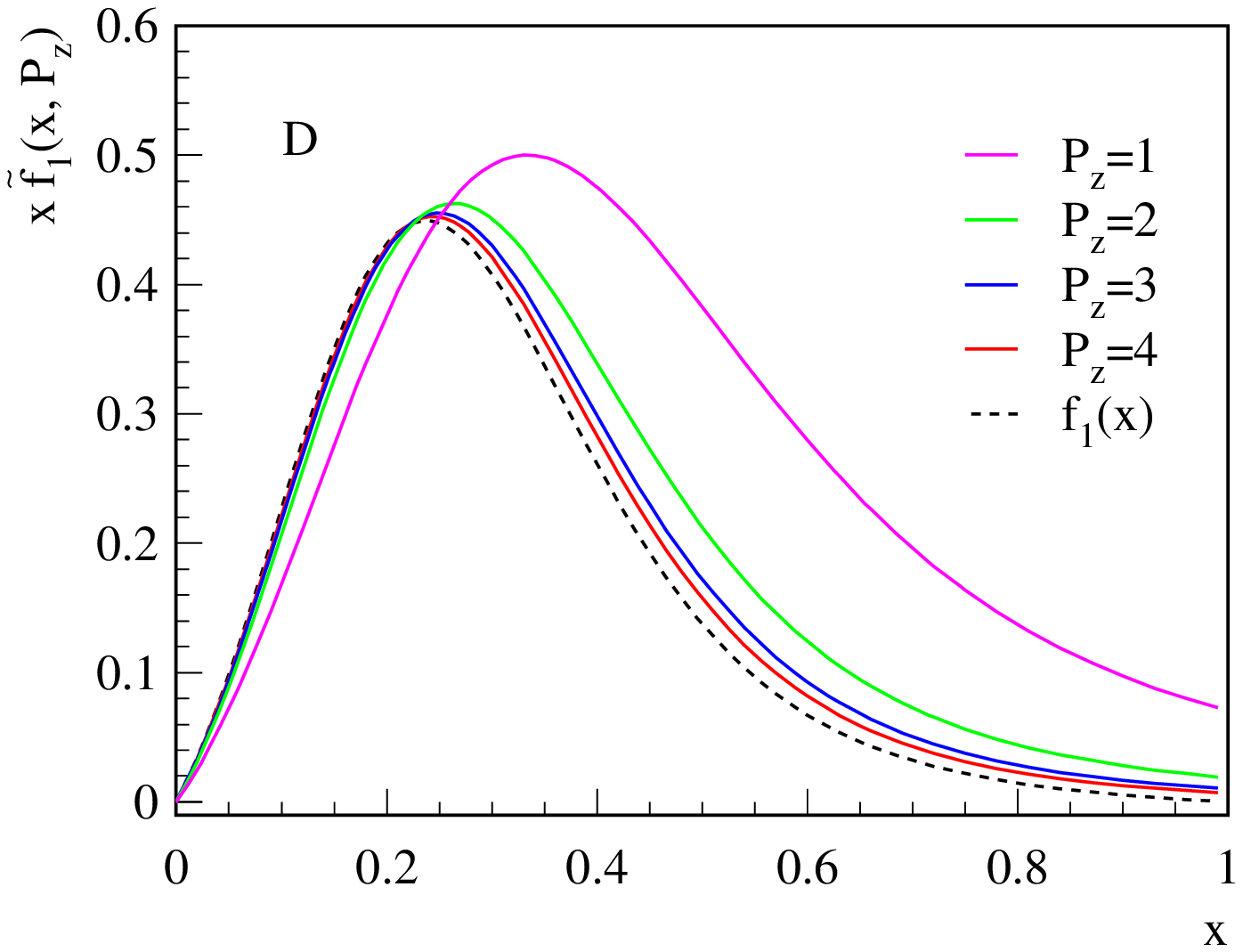, width=2.8in}
\caption{\scriptsize The unpolarized quasi-PDFs $x \tilde f_1(x, P_z)$ are plotted as a function of $x$ for $u$ (left) and $d$ (right) quark, respectively. Different lines are shown for $P_z=1$ GeV (purple), 2 GeV (green), 3 GeV (blue), and 4 GeV (red), respectively. The standard PDF $f_1(x)$ (black dashed) is also shown for comparison.}
\label{fig:f1-x}
\eef
\bef
\psfig{file=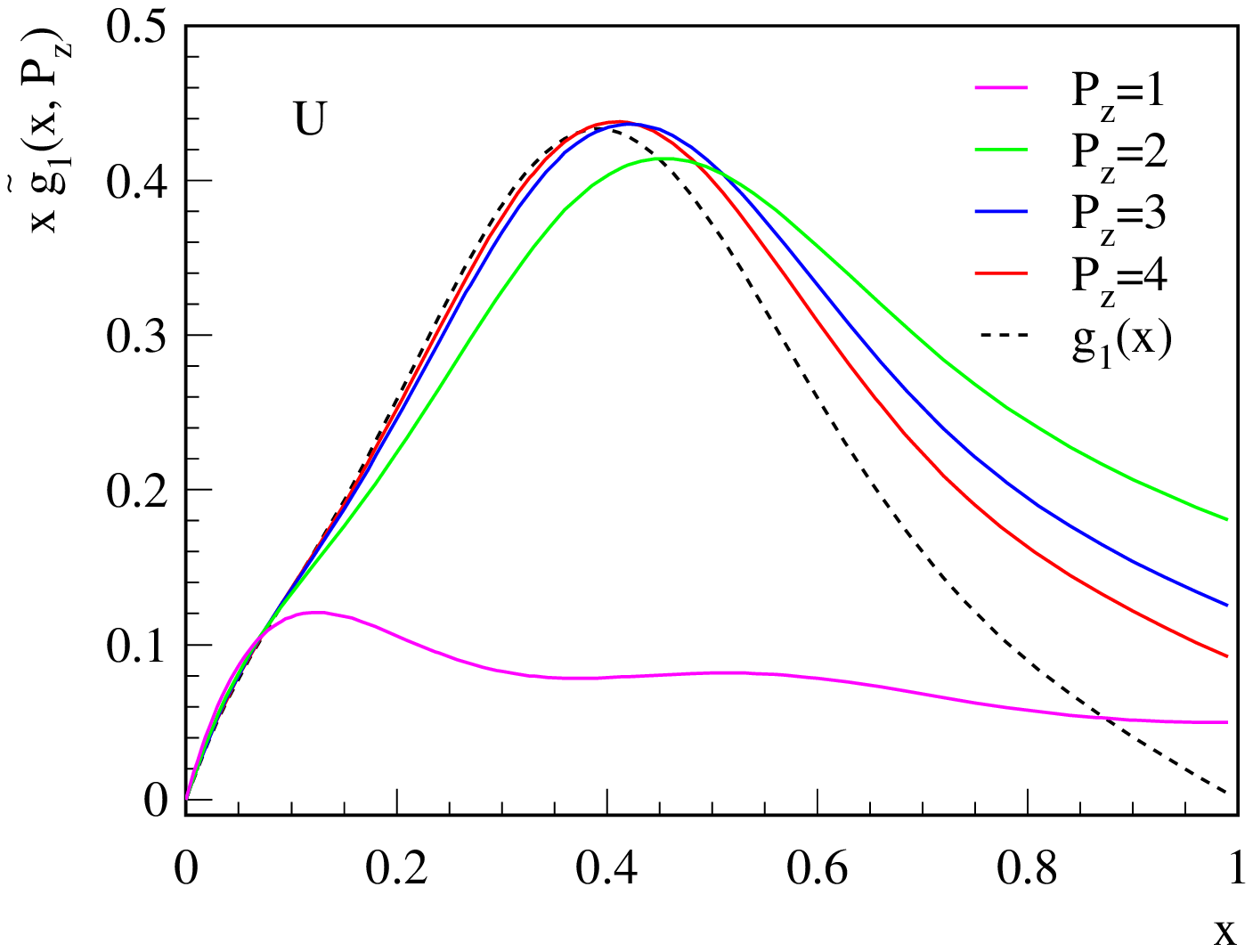, width=2.8in}
\hskip 0.4in
\psfig{file=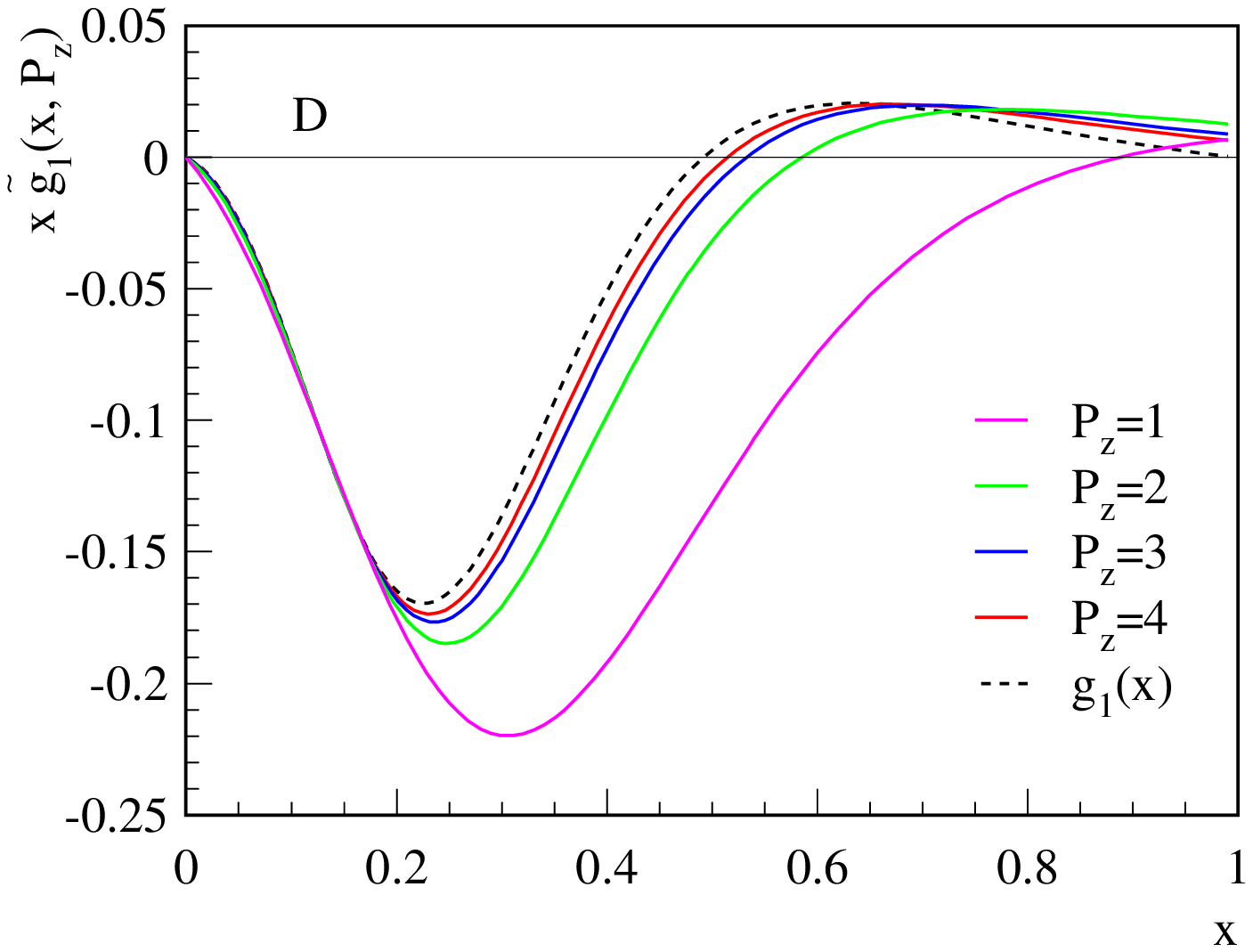, width=2.8in}
\caption{\scriptsize  The helicity quasi-PDFs $x \tilde g_1(x, P_z)$ are plotted as a function of $x$ for $u$ (left) and $d$ (right) quark, respectively. Different lines are shown for $P_z=1$ GeV (purple), 2 GeV (green), 3 GeV (blue), and 4 GeV (red), respectively. The standard helicity distribution $g_1(x)$ (black dashed) is also shown for comparison.}
\label{fig:g1-x}
\eef
\bef
\psfig{file=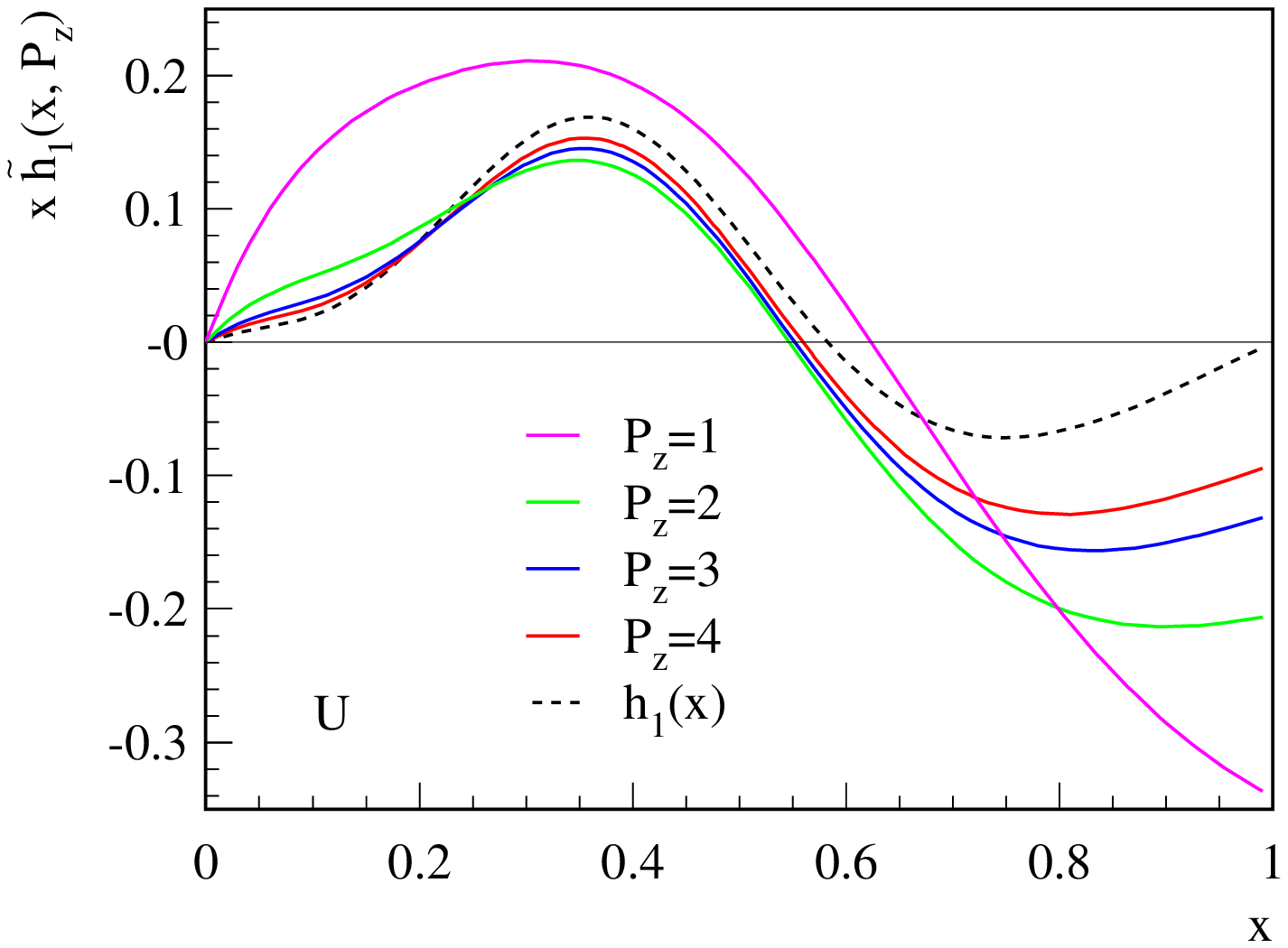, width=2.8in}
\hskip 0.4in
\psfig{file=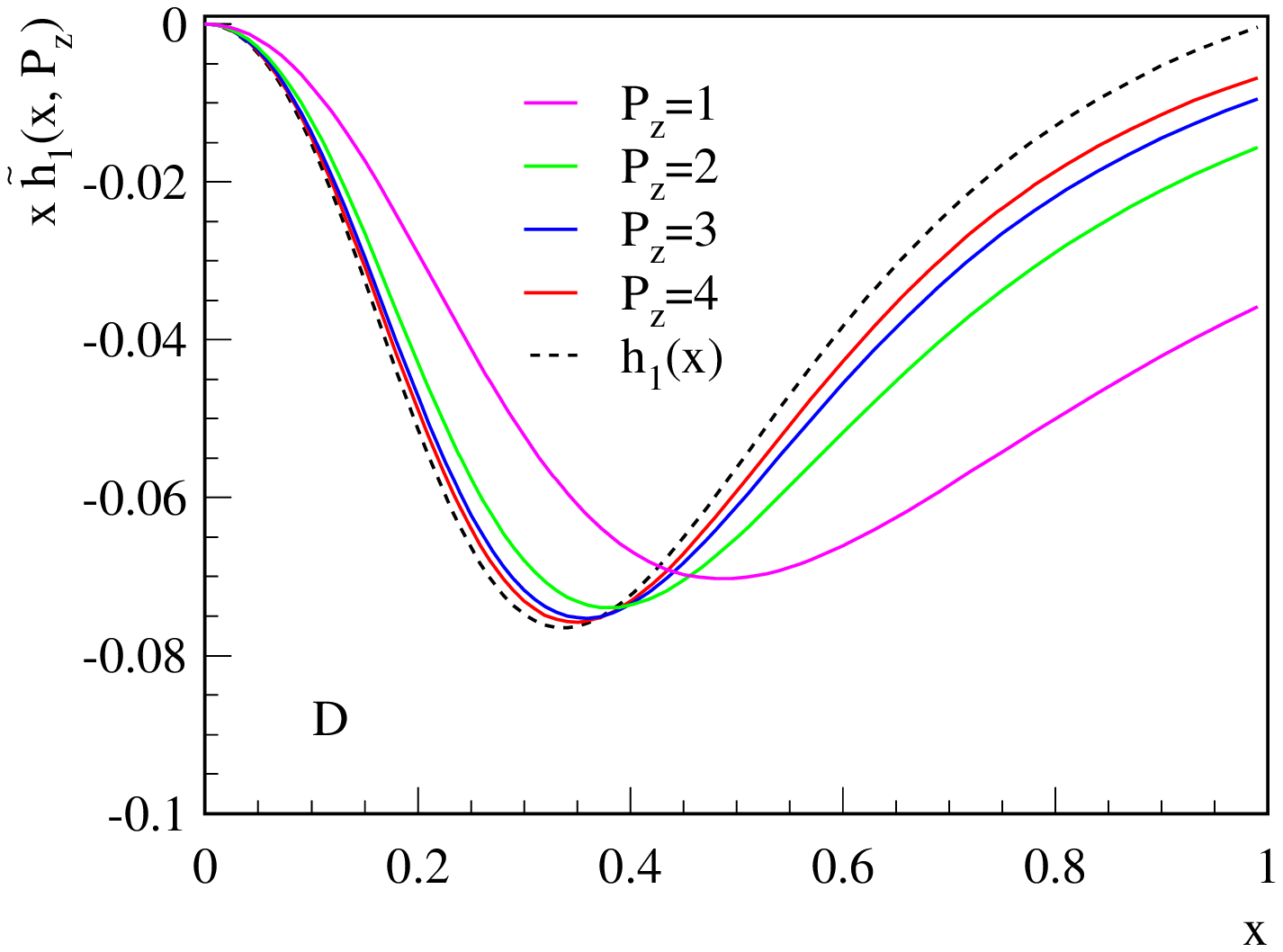, width=2.8in}
\caption{\scriptsize  The transversity quasi-PDFs $x \tilde h_1(x, P_z)$ are plotted as a function of $x$ for $u$ (left) and $d$ (right) quark, respectively. Different lines are shown for $P_z=1$ GeV (purple), 2 GeV (green), 3 GeV (blue), and 4 GeV (red), respectively. The standard helicity distribution $h_1(x)$ (blackdashed) is also shown for comparison.}
\label{fig:h1-x}
\eef

In Fig.~\ref{fig:f1-x}, we plot the quasi-unpolarized distribution $x\tilde f_1(x, P_z)$ as a function of momentum fraction $x$ for both up quark (left panel) and down quark (right panel) at different values of $P_z = 1$ GeV (purple), 2 GeV (green), 3 GeV (blue), and 4 GeV (red), respectively. For comparison, the standard unpolarized distribution $x f_1(x)$ is also shown (black dashed curve). It is important to realize that the quasi-PDFs have support for $-\infty<x<+\infty$~\cite{Ji:2013dva,Ma:2014jla,Xiong:2013bka}, and thus quasi-PDFs do not vanish for $x > 1$ at finite $P_z$. This is clearly seen in the figures: while $f_1(x)\to 0$ as $x\to 1$ for both $u$ and $d$ quarks,  at finite $P_z$, $\tilde f_1(x, P_z)$ remains finite when $x\to 1$.  It is evident that $\tilde f_1(x, P_z)$ has different behavior as compared with the standard distribution $f_1(x)$ for relatively small $P_z = 1$ GeV, as shown by the purple curves in Fig.~\ref{fig:f1-x}. However, once one increases $P_z \geq 2$ GeV, the shape of the quasi-PDFs approaches those of  the standard PDFs. 

\bef
\psfig{file=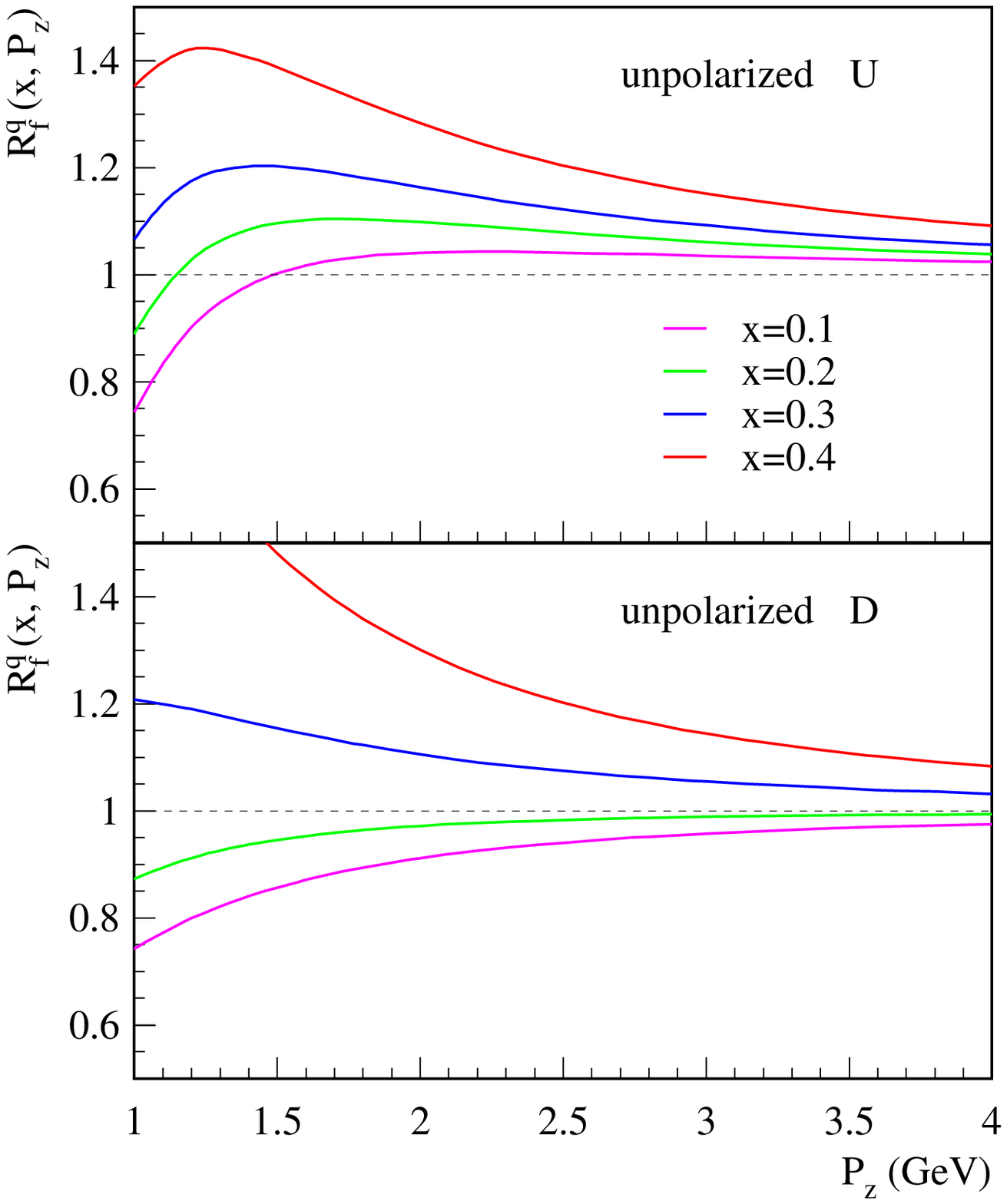, width=2.8in}
\hskip 0.4in
\psfig{file=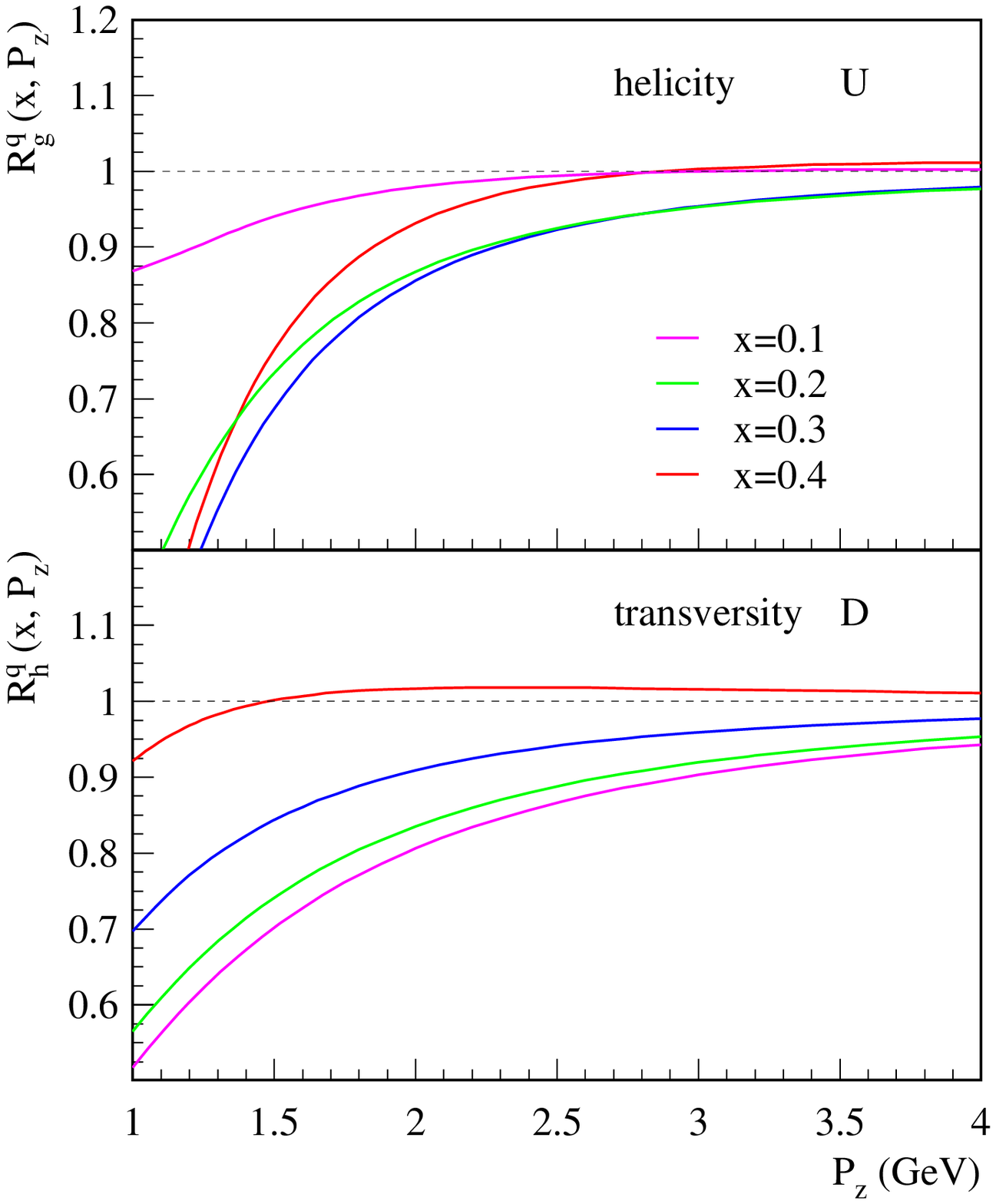, width=2.8in}
\caption{\scriptsize  The ratio $R_{f}^q(x, P_z)$ for $f_1$ (left) and $R_{g,h}^q$ for $g_1$ and $h_1$ (right) as a function of $P_z$ at different values of momentum fractions $x=0.1$ (purple), 0.2 (green), 0.3 (blue), and 0.4 (blue), respectively. On the left figure, the top (bottom) panel is for $f_1^u$ ($f_1^d$). On the right figure, the top (bottom) panel is for $g_1^u$ ($h_1^d$).}
\label{fig:ratio-ud}
\eef

In Figs.~\ref{fig:g1-x} and \ref{fig:h1-x}, 
we plot the quasi-helicity distribution $x\tilde g_1(x, P_z)$ and transversity distribution $x\tilde h_1(x, P_z)$, respectively. We find very similar features to the unpolarized case. 
 For small $P_z=1$ GeV, the quasi-PDFs are different from the standard PDFs, but again, increasing $P_z\geq 2$ GeV, they become similar to the standard PDFs. To further study the relative difference between quasi-PDFs and standard PDFs  quantitatively, we define the following ratios:
\bea
R_{f}^q(x, P_z)  =  \frac{\tilde f_1^q(x, P_z)}{f_1^q(x)},
\qquad
R_{g}^q(x, P_z)  =  \frac{\tilde g_1^q(x, P_z)}{g_1^q(x)},
\qquad
R_{h}^q(x, P_z)  =  \frac{\tilde h_1^q(x, P_z)}{h_1^q(x)},
\eea
where the subscript represents the type of the PDFs ($f_1,~g_1, ~h_1$), and the superscript denotes the quark flavor (either $u$ or $d$ quark).  In Fig.~\ref{fig:ratio-ud} (left), we present  plots of  the ratio $R_f^q(x, P_z)$ for $u$ (top panel) and $d$ (bottom panel) quark as a function of $P_z$ at different values of momentum fractions $x=0.1$ (purple), 0.2 (green), 0.3 (blue), and 0.4 (blue), respectively. In Fig.~\ref{fig:ratio-ud} (right), we present plots  for the 
$u$-quark helicity distribution $R_g^u(x, P_z)$ (top panel) and the 
$d$-quark transversity distribution $R_h^d$ (bottom panel). From these figures, it is evident that for the intermediate $0.1\lesssim x\lesssim 0.4-0.5$ all the quasi-PDFs approximate  the corresponding standard PDFs to 
within $20-30\%$ when $P_z \gtrsim 1.5 - 2$ GeV, which seems within reach of lattice QCD calculations~\cite{Lin:2014zya}. The precise values of the $P_z$ might depend on our model. However, since these features hold true for all the three collinear leading-twist quasi-PDFs $\tilde f_1$, $\tilde g_1$, and $\tilde h_1$, we expect our observation to be  generic~\footnote{We didn't present the plots here for the ratios involving $g_1^d$ and $h_1^u$. This is because with the current model parameters~\cite{Bacchetta:2008af}, $g_1^d$ and $h_1^u$ could change sign as a function of $x$, as shown in Figs.~\ref{fig:g1-x} (right) and \ref{fig:h1-x} (left). Thus the ratios $R_{g}^d$ and $R_h^u$ become unstable when $x$ approaches the node. However, we have checked that as long as $x$ stays away from the node, the ratios are similar to those in Fig.~\ref{fig:ratio-ud}.}. 

On the other hand, as we have emphasized in last section, for the very large $x\sim 1$ region, the quasi-PDFs could be quite different from standard PDFs. This has already been demonstrated in Figs.~\ref{fig:f1-x}, \ref{fig:g1-x}, and \ref{fig:h1-x}, where the quasi-PDFs are still finite but the standard PDFs all vanish when $x\to 1$. Let us further make this point. In Fig.~\ref{fig:ratio-largex}, we plot the ratio $R^q(x, P_z)$ at large $x=0.7$ as a function of $P_z$ for $f_1^u$ (red), $f_1^d$ (blue), $g_1^u$ (green), and $h_1^d$ (purple), respectively. One can see that at $P_z \sim 1-2$ GeV, the ratio can be as large as $6-7$; that is,  in the large $x$ kinematics regime, the
 quasi-PDFs are quite different from the standard PDFs. In this 
one has to go to very large $P_z > 4$ GeV at least to obtain a good approximation to the standard PDFs. 
\bef
\psfig{file=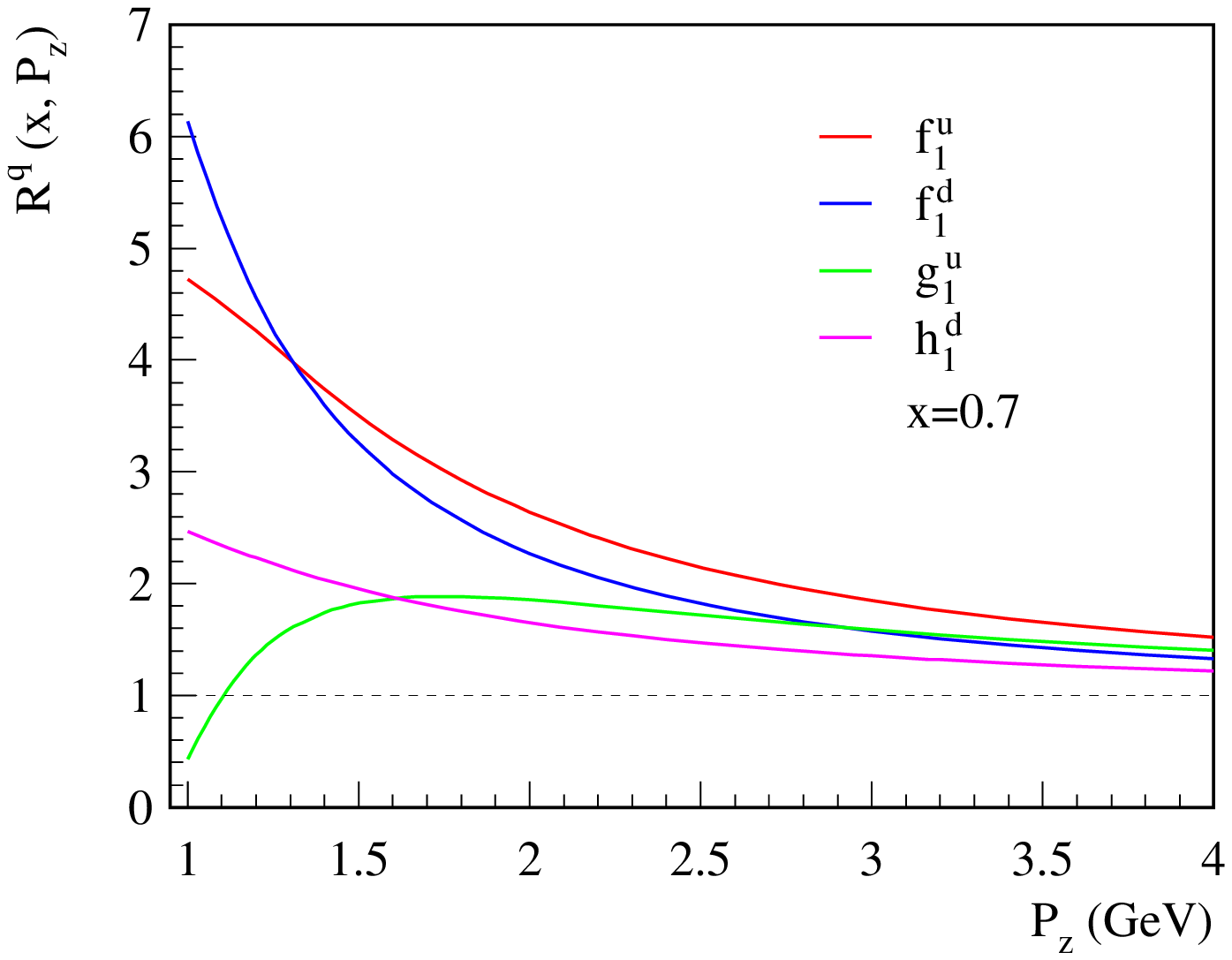, width=2.8in}
\caption{\scriptsize The ratio $R^q(x, P_z)$ at large $x=0.7$ as a function of $P_z$ for $f_1^u$ (red), $f_1^d$ (blue), $g_1^u$ (green), and $h_1^d$ (purple), respectively.}
\label{fig:ratio-largex}
\eef

After the discussion on the individual quasi-PDFs, let us study the relation between them. As we have mentioned already in the last section, the Soffer inequality relates the three leading-twist collinear PDFs $f_1,~g_1$,  and $h_1$ as in Eq.~\eqref{eq:soffer}. To test such an inequality for both standard PDFs and quasi-PDFs, let us define the following quantities:
\bea
\tilde S^q(x, P_z) & = \frac{1}{2}\left(f_1^{q}(x, P_z) + g_1^{q}(x, P_z)\right) - \left|h_1^{q}(x, P_z)\right|,
\\
S^q(x) & = \frac{1}{2}\left(f_1^{q}(x) + g_1^{q}(x)\right) - \left|h_1^{q}(x)\right|.
\eea
The Soffer bound holds for the standard PDFs, thus we have, 
\bea
S^q(x) \geq 0,
\eea
as has already been demonstrated in the spectator diquark model in the last section. Now let us test whether Soffer bound is satisfied for the quasi-PDFs, in other words, whether
\bea
\tilde S^q(x, P_z)  \stackrel{?}{\geq} 0,
\eea
for any $x$ and $P_z$.
\bef
\psfig{file=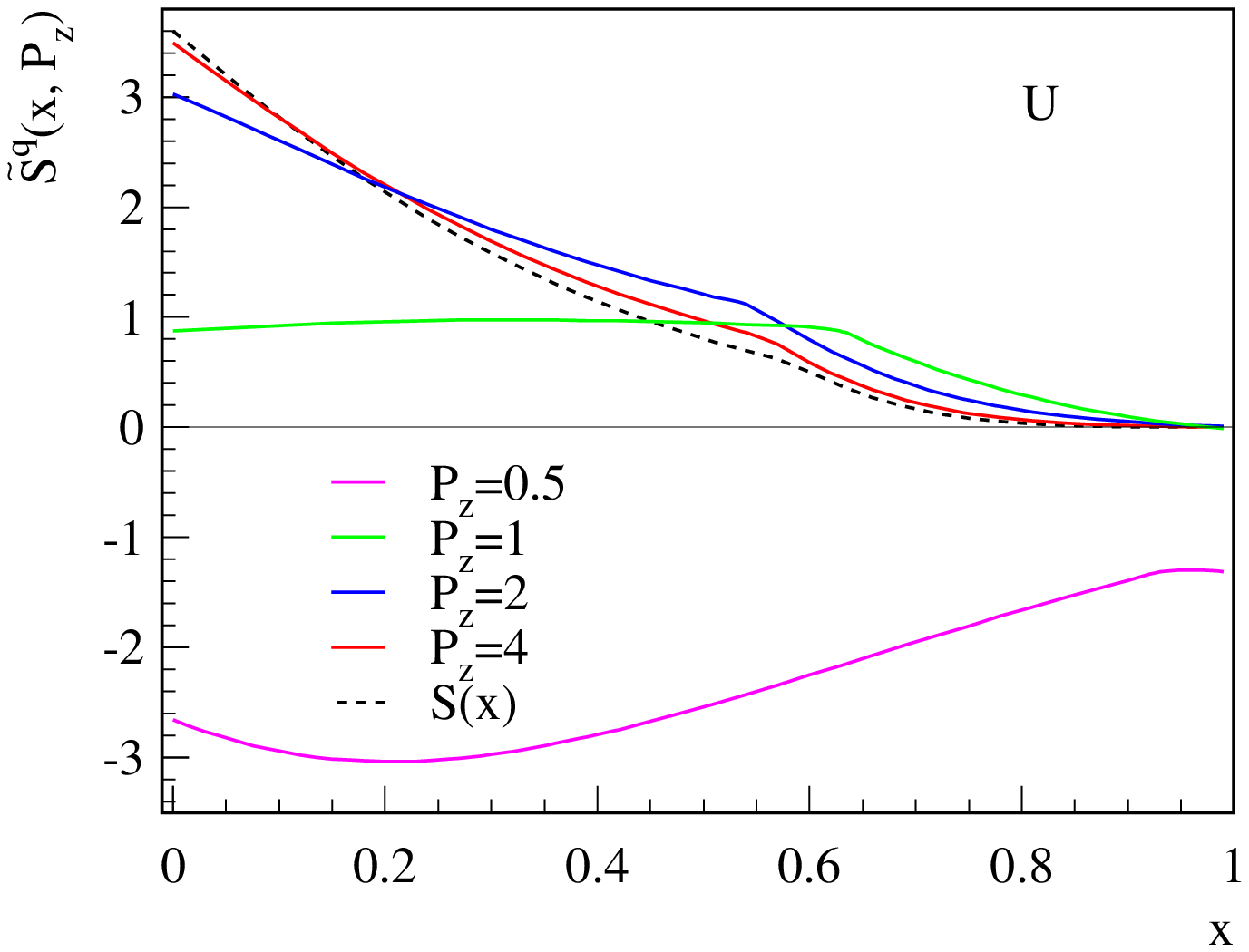, width=2.8in}
\hskip 0.4in
\psfig{file=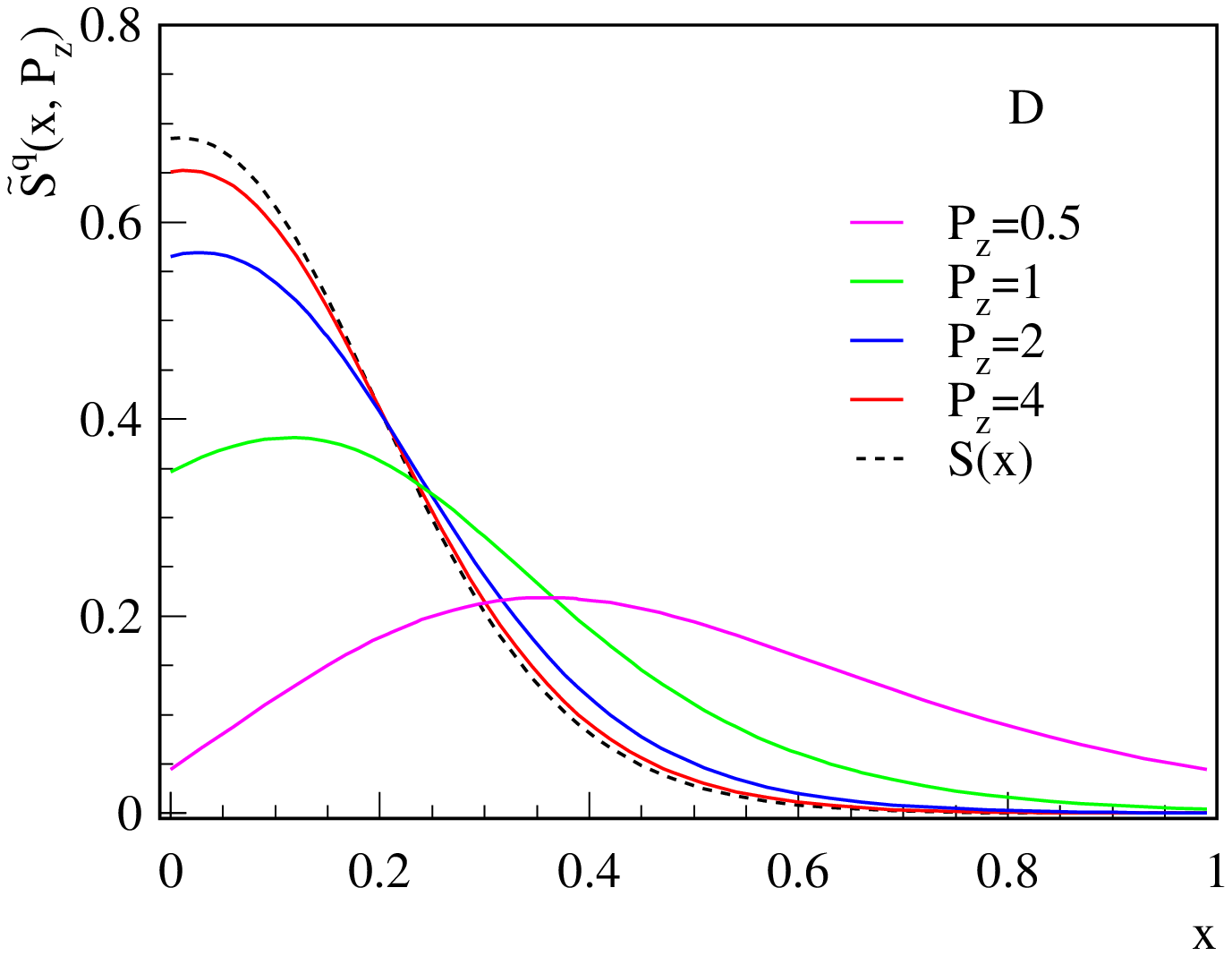, width=2.8in}
\caption{\scriptsize  The function $\tilde S^q(x, P_z)$ is plotted versus $x$ for $u$ (left) and $d$ (right) quark. Different lines are shown for $P_z=0.5$ GeV (purple), 1 GeV (green), 2 GeV (blue),  4 GeV (red), respectively. The function $S^q(x)$ (black dashed) is also shown for comparison.}
\label{fig:soffer}
\eef
In Fig.~\ref{fig:soffer}, $\tilde S^q(x, P_z)$ is plotted versus $x$ for $u$ (left) and $d$ (right) quark at different values of  $P_z$, $0.5$ GeV (purple), 1 GeV (green), 2 GeV (blue), and 4 GeV (red), respectively. The function $S^q(x)$ (black dashed) for the standard PDFs is also shown for comparison. As one can see clearly from the black dashed curves, the Soffer bound is indeed satisfied for the standard PDFs for both $u$ and $d$ quarks. At the same time, within our spectator diquark model, as shown in the right panel of Fig.~\ref{fig:soffer}, for all the selected $P_z$ values, $\tilde S^q(x, P_z) \geq 0$ for the $d$ quark, 
that is,  the Soffer bound appears  to be satisfied for the $d$-quark quasi-PDFs. On the other hand, 
as shown in the left panel of Fig.~\ref{fig:soffer} for the $u$ quark, even though $\tilde S^q(x, P_z) \geq 0$ for $P_z=1,~2$, and 4 GeV,  for $P_z = 0.5$ GeV, $\tilde S^q(x, P_z) < 0$ for the entire plotted $0\leq x\leq 1$ region. In other words, the Soffer bound breaks down for relatively small $P_z$ values for the $u$ quark. What this tells us for  the usual lattice QCD simulations is that while the standard PDFs might still satisfy the positivity bounds, such as Soffer bound on the lattice~\cite{Diehl:2005ev}, these positivity bounds in general do not hold for quasi-PDFs, and, thus, one should avoid using them in lattice simulations.

\vspace{-0.3cm}
\section{Summary}
\vspace{-0.3cm}
We  studied the quasi-parton distribution functions (PDFs) and standard PDFs consistently within the framework of the spectator diquark model. Our work aims to answer the question to what values of the proton momentum $P_z$ the quasi-PDFs are good approximations for the standard PDFs. We took into account both the scalar diquark and axial-vector diquark contributions and generated  all the three leading-twist collinear PDFs (the unpolarized distribution~$f_1$, 
the helicity distribution~$g_1$, and the transversity distribution~$h_1$) for both up and down quarks. Using the model parameters which lead to a reasonable description of the standard PDFs~$f_1^{u,d}(x)$ and~$g_1^{u,d}(x)$, consistent with those extracted from the global analysis (see~\cite{Bacchetta:2008af}), we presented  numerical studies for all quasi-PDFs. We found that for intermediate $0.1\lesssim x \lesssim 0.4-0.5$, the quasi-PDFs are good approximations for the corresponding standard PDFs  when the proton momentum $P_z \gtrsim 1.5-2$ GeV. Such kinematics appears feasible for 
 lattice QCD calculations. However, in the large $x\sim 1$ region, a much larger $P_z > 4$~GeV  is necessary to obtain a similar accuracy of the approximation. By studying the Soffer positivity bound we found that the positivity bounds do not hold in general for the quasi-PDFs. Our study provides useful guidance for the lattice QCD calculations regarding the proton boost and accuracy of the quasi-PDFs approximation.

\vspace{ -0.3cm} 
\section*{Acknowledgements}
\vspace{ -0.3cm} 
This work is supported by the U.S. Department of Energy under Contract Nos. DE-FG02-07ER41460 (L.G.) and DE-AC02-05CH11231 (Z.K., I.V. and H.X.), and in part by the LDRD program at LANL.

\bibliographystyle{h-physrev5}   
\bibliography{biblio}

\end{document}